\documentclass[%
 reprint,
superscriptaddress,
 amsmath,amssymb,
 aps,
float]{revtex4-1}
\usepackage{graphicx}
\usepackage{epstopdf}
\usepackage{dcolumn}
\usepackage{bm}
\usepackage[caption=false]{subfig}

\begin{document}

\title{Gaseous $^3$He Nuclear Magnetic Resonance Probe for Cryogenic Environments}

\author{X. Fan}
\affiliation{Department of Physics, Harvard University, Cambridge, Massachusetts 02138, USA}
 \affiliation{Center for Fundamental Physics, Northwestern University, Evanston, Illinois 60208, USA}
 \author{S. E. Fayer}
 \affiliation{Center for Fundamental Physics, Northwestern University, Evanston, Illinois 60208, USA}
 \author{G. Gabrielse}
 \email{gerald.gabrielse@northwestern.edu}
 \affiliation{Center for Fundamental Physics, Northwestern University, Evanston, Illinois 60208, USA}
\date{\today}

\begin{abstract}
Normal nuclear magnetic resonance (NMR) probes cannot be used to make high frequency resolution measurements in a cryogenic environment because they lose their frequency resolution when the liquid sample in the probe freezes.  A gaseous $^3$He NMR probe, designed and constructed to work naturally in such cryogenic environments, is demonstrated at 4.2 K and 5.3 Tesla to have a frequency resolution better than 0.4 part per billion.  As a demonstration of its usefulness, the cryogenic probe is used to shim a superconducting solenoid with a cryogenic interior to produce a magnetic field with a high spatial homogeneity, and to measure the magnetic field stability.
\end{abstract}

\maketitle










\section{Introduction}

Small liquid samples (e.g.\ water or acetone) have very narrow NMR resonance linewidths (e.g. \cite{PRIGL1996118}), so they are frequently used to characterize the magnetic field produced by superconducting solenoids.  Unfortunately, such probes cannot be used with solenoid systems that have only low temperature interior volumes because the liquid samples would freeze and lose their narrow linewidth.   Heaters or dewars installed to prevent freezing are not ideal as they can perturb the magnetic field being characterized.         

We are particularly interested in characterizing the magnetic field within the cryogenic volume used to measure the electron and positron magnetic moments \cite{HarvardMagneticMoment2011,DehmeltMagneticMoment}.  The moment of a single isolated particle is deduced from spin and cyclotron frequencies, each of which is measured to extremely high accuracy and proportional to the magnetic field.  A spatial homogeneity on the order of a part in $10^{8}$ over a centimeter and a time stability on the order of a part in $10^{10}$ per hour are required within a 4.2 K bath, normally containing a 0.1 K Penning trap \cite{EfficientPositronAccumulation}. The electron and positron moments, the most accurately measured properties of elementary particles, are the most precise predictions of the standard model of particle physics \cite{TowardsElectronPositronMoments2019} and provide the most sensitive direct test of its fundamental charge conjugation, parity and time reversal (CPT) symmetry invariance with leptons.

\begin{figure*}
\centering
    \begin{minipage}{3.375in}
        \centering
    \includegraphics[width=3.in]{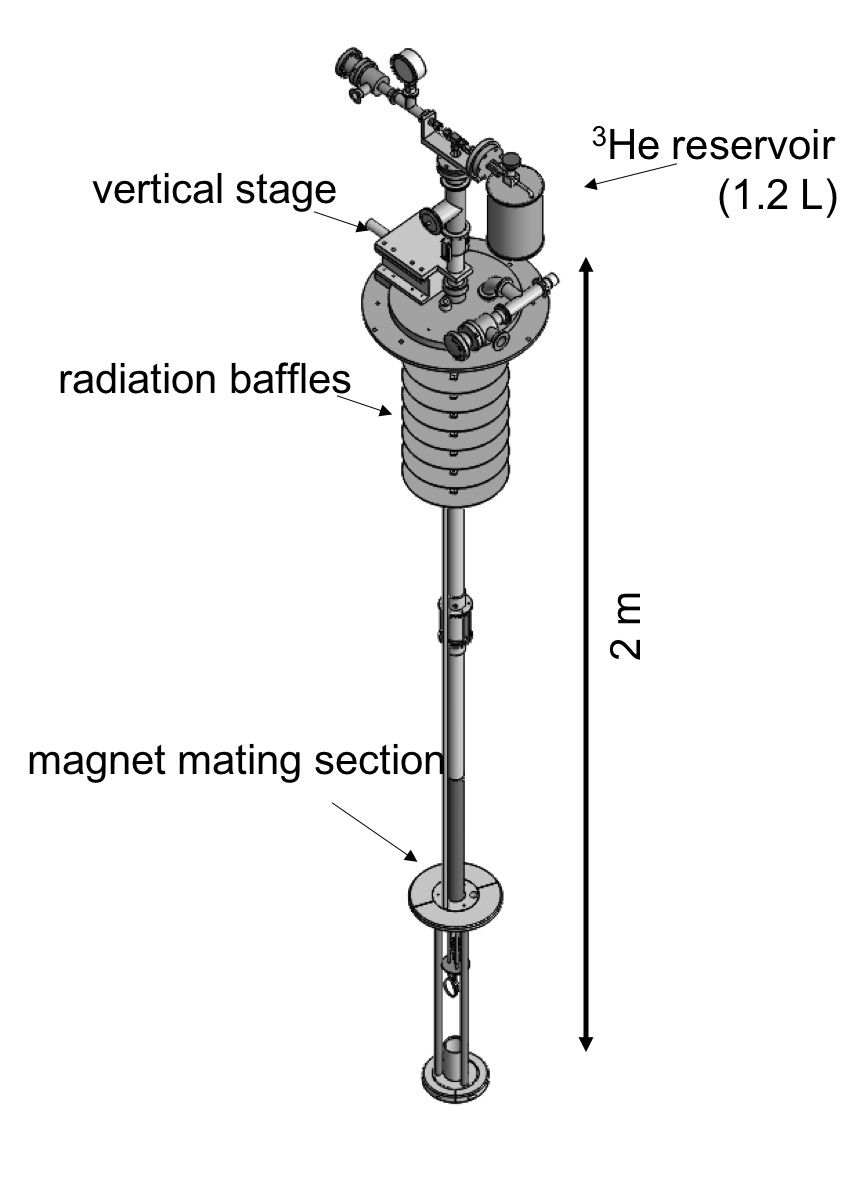}
    \end{minipage}\hfill
    \begin{minipage}{3.375in}
        \centering
        \includegraphics[width=2.5in]{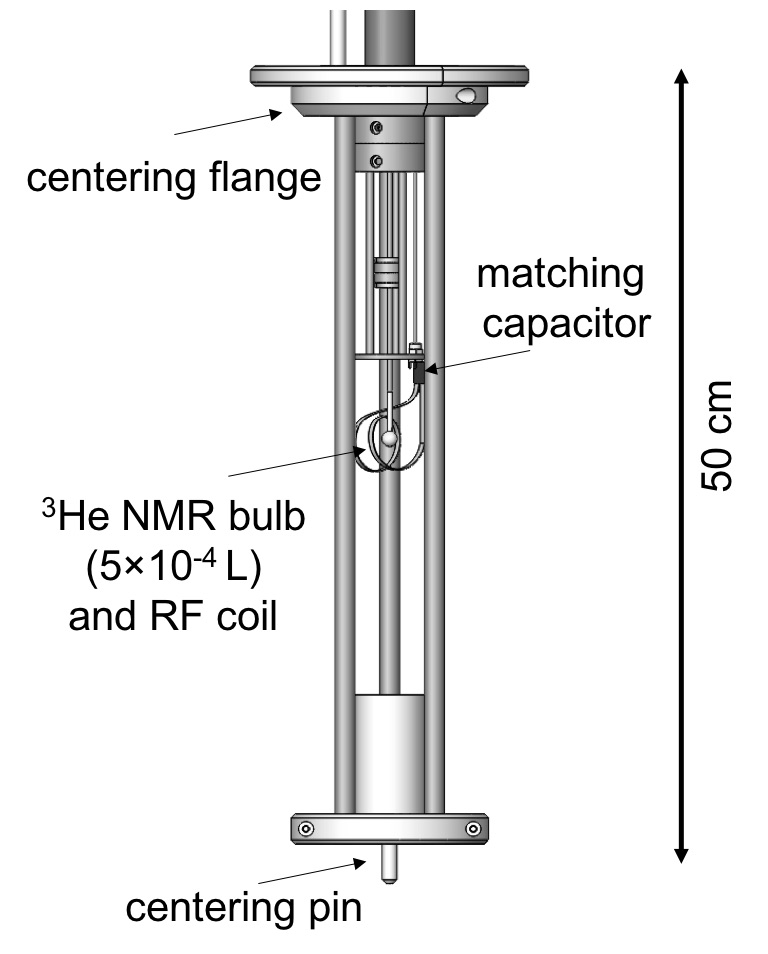}
    \end{minipage}
	    \caption{(left) Overview of an NMR probe and support, designed to align the bulb with the axis of the superconducting solenoid being used for positron and electron measurements. The large room temperature reservoir is connected to the NMR bulb via a capillary tube to ensure high density of $^3$He target. The vertical translation stage makes it possible to raise and lower the NMR sample bulb. (right) Expanded view of the NMR sample bulb and its pickup coil. The centering flange and pin ensure that the bulb is radially aligned with the magnet. The RF coil is made of 99.9999 $\%$ purity copper and is loosely wound around the NMR bulb to minimize the residual magnetism. The matching capacitor is attached on a board near the bulb, see also Fig. \ref{fig:NMRCircuit}. The magnetism of all of the components has been carefully checked.}
    \label{fig:NmrProbeOverview}
\end{figure*}

The $^3$He gas NMR probe demonstrated here (Fig.~\ref{fig:NmrProbeOverview}) works naturally at cryogenic temperatures.  It is demonstrated at 4.2 K but should work at lower temperatures. $^3$He is used because it has a substantial nuclear moment and it remains a gas at low temperatures. The major challenges are that many fewer spins are typically available to make a NMR signal in a gas compared to a liquid in the same volume, motional narrowing needs to be suppressed to measure the inhomogeneity correctly, and $^3$He is extremely expensive. The polarization fraction is much larger at low temperature, but that still does not compensate the lower number of spins.  In order to greatly increase the number of spins without increasing the sample pressure to a dangerous level, $^3$He is kept in a closed system that has a large reservoir in addition to a small NMR bulb. $^3$He moves from the room temperature reservoir to the small probe volume when the latter is cooled to 4.2 K with the pressure inside the bulb remaining approximately constant.  The result is a signal from a 4.2 K gas sample that is comparable to that from a room temperature water sample. The 1 atm pressure at 4.2 K makes the diffusion constant is low enough to suppress motional narrowing. As a result, the inhomogeneity of the field can be directly deduced from the transverse decay constant of the spin precession signal. The crucial NMR relaxation times $T_1$, $T_2$, and $T_2^*$ for $^3$He at 4.2 K are deduced from free precession NMR signals and spin echoes. 

We illustrate the usefulness of the gaseous $^3$He NMR probe by shimming and characterizing a new superconducting solenoid system with a cold bore.  The 5.3 T magnetic field is mostly produced by a persistent current in a large superconducting solenoid.  The persistent currents injected into 11 superconducting shim coils are adjusted to narrow the frequency width of the NMR signal. The narrow NMR resonance is then used to precisely measure the stability of the magnetic field.  

Note that in non-cryogenic contexts, $^3$He NMR precession signals have been observed using room temperature gas cells and much smaller magnetic fields \cite{PhysRevC.36.2244}. Key to such measurements was using lasers to attain nearly complete initial polarization \cite{Gentile2017}, much larger than is produced for an NMR sample in thermal equilibrium. This could be added to cryogenic NMR probe but it would add considerable complexity and was not required for the demonstrated precision.

\section{Spin alignment}

The NMR signal is proportional to the size of the net magnetic moment: 
\begin{equation}
    M=N\mu\tanh{\frac{\mu B}{k_B T}} \approx N\mu\frac{\mu B}{k_B T},
    \label{eq:TotalMagneticMoment}
\end{equation}
 where $N$ is the number of spins in thermal equilibrium at temperature $T$, $\mu$ is the magnetic moment of each spin and $k_B$ is the Boltzmann constant.  The hyperbolic tangent factor is the net fraction of the spins that are thermally aligned. The approximation to the right in Eq.~(\ref{eq:TotalMagneticMoment}) suffices in all cases considered here.
 
Since a room temperature water sample in a 1 cm diameter spherical cell produces a large enough NMR signal to be useful, we compare the size of the water moment to that for $^3$He gas in the same sample volume in  Table \ref{table:CellParameters}.  The magnetic moments, $\mu$, are given in nuclear magnetons $\mu_N$.  Since the $^3$He moment is 76\% that of water, largest differences between the net moments come from the differing numbers of spins in the cell, $N$, and polarization fractions, $\mu B/(k_B T)$.  

\begin{table}[h!]
\begin{tabular}{l|c|c|c|c}
                & $\mu/\mu_N$ &        $N$               &  $\mu B/(k_B T)$   &    $M/\mu_N$    \\
\hline
H$_2$O          & 2.8        & $3.5\times10^{22}$   &  $1.8\times 10^{-5}$  & $1.8\times10^{18}$ \\
$^3$He (only gas cell)          & 2.1         & $1.3\times10^{19}$      & $9.8\times 10^{-4}$ & $2.7\times 10^{16}$ \\
$^3$He (with reservoir)          & 2.1         & $9.1\times10^{20}$   & $9.8\times 10^{-4}$ & $1.9\times 10^{18}$ \\
\end{tabular}
\caption{Comparison of NMR samples discussed at $B=5.3$ Tesla. The values for H$_2$O are calculated at 300 K, and those for $^3$He are calculated at 4.2 K. See text for details.}
\label{table:CellParameters}
\end{table}

An atmosphere of $^3$He in the same sample cell volume, after being cooled from 300 K to 4.2 K, results in $\sim 2700$ times fewer spins than for the water sample (first $^3$He line in the table). Even though the polarization factor increases by a factor of 54, the net magnetic moment (and hence the size of the NMR signal) is only 1.5\% that of the water sample.  

Increasing the room temperature pressure inside a sealed bulb to match the water signal would require 60 atmospheres of pressure in the bulb at room temperature. Instead, we connect the 0.5 cm$^3$ glass bulb through a capillary to a much larger (1.2 liter) reservoir volume that stays at room temperature.  Gas atoms move from the reservoir into the bulb to keep approximately 1 atmosphere of pressure in the bulb as it cools to 4.2 K.  The last line in the table shows that the number of nuclear spins in the bulb is still 38 times smaller than for the water sample. However, as the polarization fraction is 54 times larger, the net result (when the slightly different nuclear moments are also factored in) is that the magnetic moment of the gas sample is 1.1 times that of the water sample.  The NMR signal size is thus 10\% bigger than a room temperature water sample would produce in the same volume.

This high density condition also suppresses the effect of motional narrowing \cite{PhysRevA.37.2877}. At 4.2 K and 1 atmospheric pressure, the diffusion coefficient of $^3$He is as small as $D\approx 4\times10^{-7}$ m$^2$/s, see Sec.\ref{sec:timeconstants} and \cite{PhysRevA.39.6170}. With this slow diffusion rate, we are in the limit $D\rightarrow 0$ and motional narrowing does not occur. This ensures that the measured $T_2^*$ reflects the inhomogeneity over the volume directly.

Figure \ref{fig:NmrProbeOverview} shows the gaseous $^3$He probe used in this demonstration.  It was constructed to characterize a new superconducting solenoid system intended to make electron and positron magnetic moment measurements.  The 1 cm diameter glass bulb (Type I, Class A borosilicate glass, 529-A-12 Wilmad-LabGlass) is produced for NMR use. We measured it to have negligible magnetism at the level discussed here. All other probe components were measured to have minimal magnetism and placed as far as possible from the bulb to avoid other magnetic perturbations.  Special care was taken with the alignment parts in the magnet mating section, which were fabricated from only pure copper, aluminum, molybdenum and titanium. The RF coil near the bulb is made from a 99.9999 $\%$ pure thin copper foil, and is loosely wound around the bulb. The centering flange and centering pin ensure the radial alignment with our magnet. The $^3$He line is mechanically fixed to the vertical stage at the hat, and can be moved inside the magnet bore.  We can rotate NMR bulb, capillary line, and the electronics inside the magnet bore as one check of the residual magnetism of the probe.

\section{NMR Spin Precession Signal}
\label{sec:spinprecession}

The NMR probe is demonstrated using a $B=5.3~\textrm{T}$ magnetic field in the $\hat{\mathbf{z}}$ direction, generating a $^3$He spin precession frequency of $\omega_0/2\pi=172.3$ MHz. The homogeneity of the field in this direction is optimized by changing the currents on 11 shim coils. Three of these produce linear gradient fields that go primarily as x, y and z, and five shims produce quadratic gradients that go mostly as z$^2$, xy, yz, xz, and x$^2-$y$^2$.  The remaining three shims produce fields that go primarily as z$^3$, z$^2$x, and z$^2$y. 

\begin{figure}
\centering
\subfloat{\includegraphics[width=\the\columnwidth]{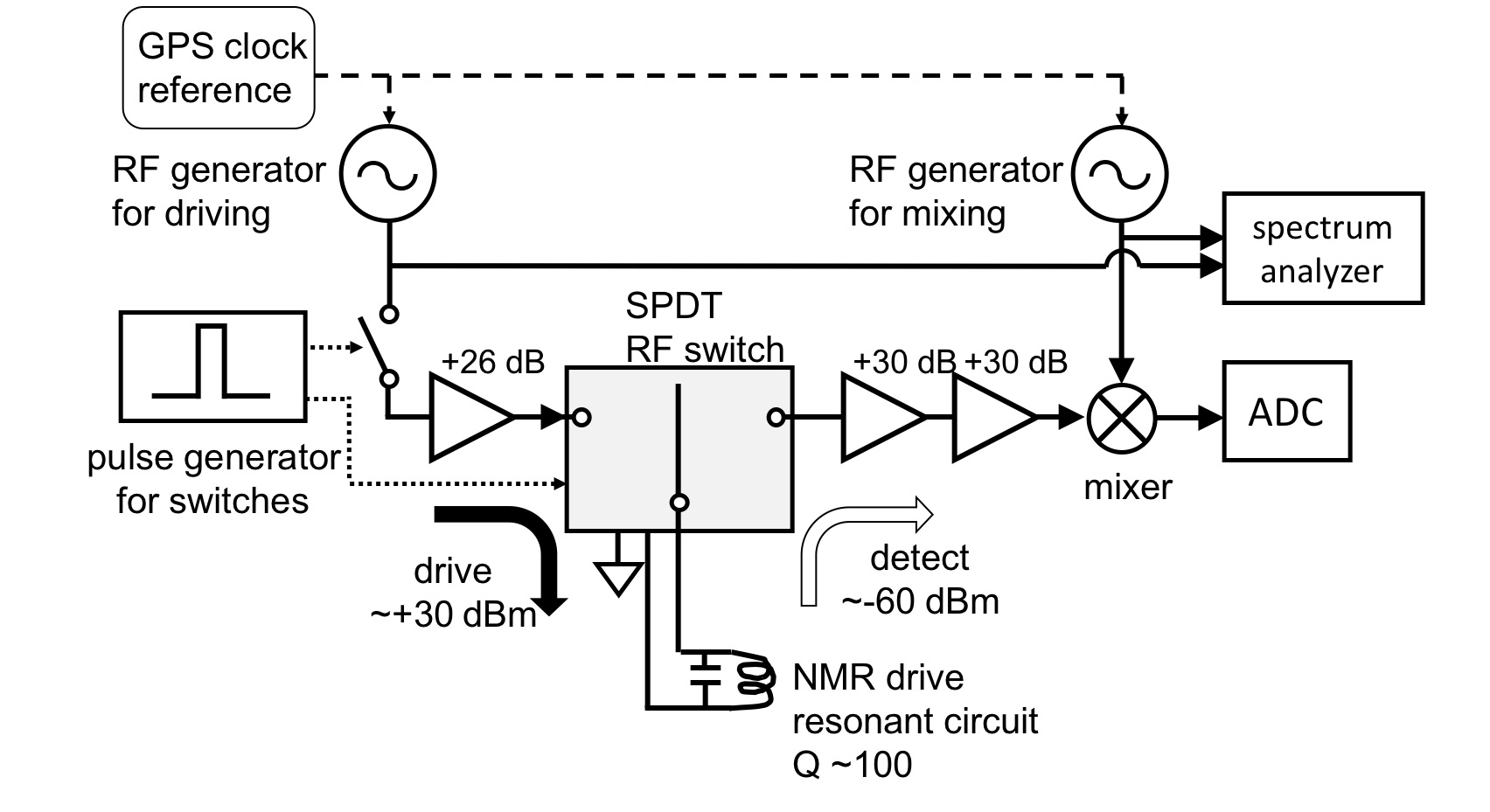}}
\hfill
\subfloat{\includegraphics[width=\the\columnwidth]{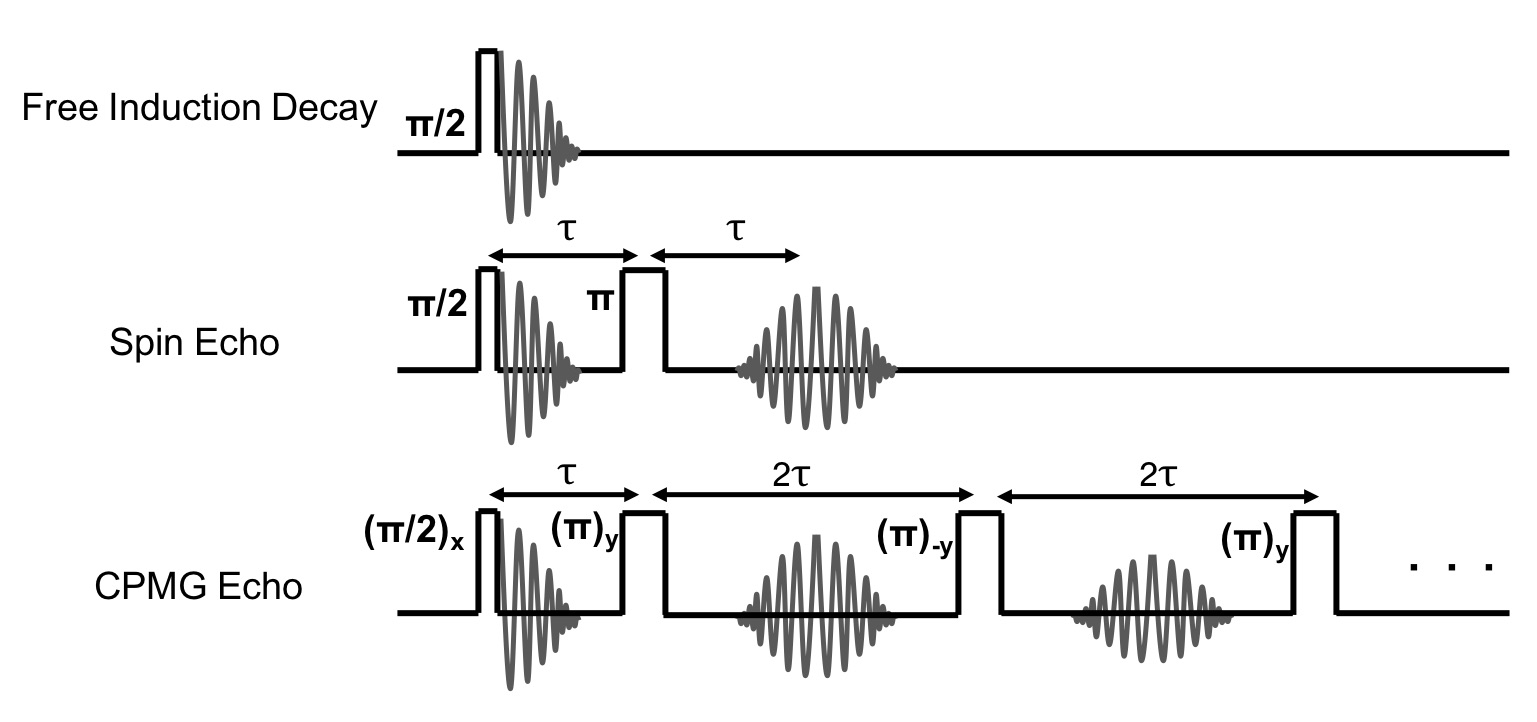}}
\caption{(Top) Circuit used to drive and detect the $172.3$ MHz NMR signal.  (Bottom) The drive pulse sequences for free induction decay (Sec.\ref{sec:spinprecession}) and for both simple and CPMG spin echos (Sec.\ref{sec:timeconstants}).}   \label{fig:NMRCircuit}
\end{figure}

The circuit used to produce and detect the NMR signal from the $^3$He is a simple switching circuit shown in Fig.~\ref{fig:NMRCircuit}. Two RF frequency generators referencing a GPS clock signal are used. One drives the $^3$He spin at the resonant frequency, and the other is used to mix down the NMR signal down to roughly 1.5 kHz. Then the signal is recorded by an Analog-to-Digital converter (ADC). Both frequencies are monitored by a spectrum analyzer. A pulse-controlled single pole, double throw (SPDT) RF switch is used to switch driving and detection side. Since the signal isolation of the SPDT switch is not enough, another RF switch is used in the driving side to suppress direct feedthrough. Three RF amplifiers are used to drive and detect the NMR frequency. A matching capacitor is mounted near the NMR RF coil in the liquid helium to form a resonant circuit with a Q-factor of about 100, which increases both driving and detection efficiency. The three applied pulse sequences are also shown in Fig.~\ref{fig:NMRCircuit}, details of each of these are given in the following sections. 

\begin{figure}
    \centering
    \includegraphics[width=\the\columnwidth]{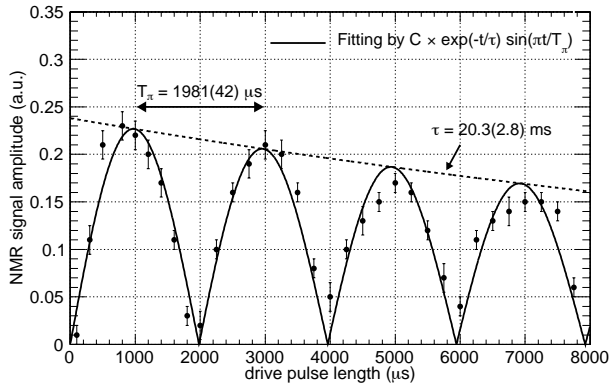}
    \caption{The magnitude of the NMR precession signal depends upon the strength and duration of the drive pulse, the latter being varied here. Each peak in the graph correspond to $\pi/2$, $3\pi/2$, $5\pi/2$, and so on, respectively. The decay constant is due to the inhomogeneity of the RF drive intensity. By taking this scan, we can measure the pi-pulse drive length $T_{\pi}$. }  \label{fig:PiHalfMeasurement}
\end{figure}

In a thermal equilibrium at temperature $T$, the population in the lower of the two spin states is slightly less, as described in Eq.~\ref{eq:TotalMagneticMoment}. The $^3$He bulb has the net magnetic moment given in the last line of Table \ref{table:CellParameters}.  A nearly resonant drive pulse tips the resulting magnetic moment vector by an angle of $\pi/2$, as is typical in pulsed NMR measurements\cite{PhysRev.80.580}. The size of the NMR signal depends upon the length of the drive pulse and on the drive intensity, as well as on the net magnetic moment.  Fig.~\ref{fig:PiHalfMeasurement} shows how the initial signal size varies as a function of the length of the drive pulse. 

\begin{figure}
		\centering
	    \includegraphics[width=\the\columnwidth]{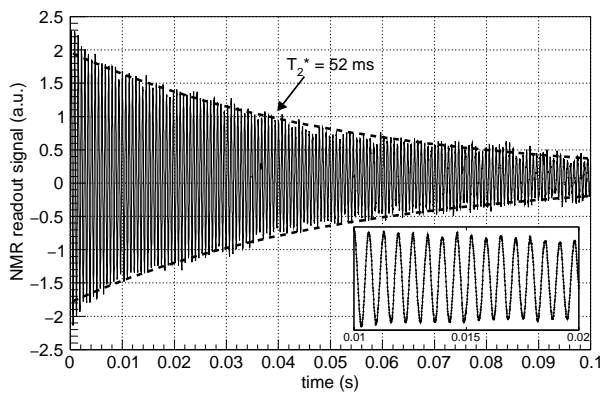}
    \caption{The NMR spin precession signal from the $^3$He nuclei at 5.3 T, mixed down from 172.3 MHz to 1.5 kHz, decays with a time constant $T_2^*$ = 52 ms in this example. The dashed lines represent the exponential decay with $T_2^*$ obtained by a fitting. The inset shows an expanded view of the same plot from 0.01 s to 0.02 s. A clear sinusoidal oscillation signal is observed.}  \label{fig:NmrFID}
\end{figure}

\begin{figure}
        \centering
    	\includegraphics[width=\the\columnwidth]{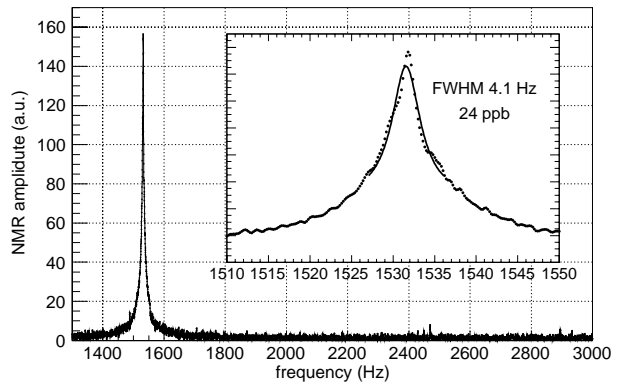}
    \caption{Fourier transform of the NMR spin precession signal shown in Fig.~\ref{fig:NmrFID}. The signal-to-noise ratio is about 160. The inset shows an expanded view of the NMR peak. A Lorentzian fitting of this peak gives a FWHM of 4.1 Hz, which corresponds to 24 ppb relative inhomogeneity.}  \label{fig:NmrSignalFFT}
\end{figure}

The polarization, now tipped perpendicular to the magnetic field direction, rotates at the NMR angular frequency $\omega_0$ around the magnetic field direction. The changing flux through the pickup coils induces an electromagnetic field across the coil, which is detected.  Fig.~\ref{fig:NmrFID} shows how the free induction decay (FID) signal at 5.3 T, mixed down to about 1.5 kHz, decays with a time constant $T_2^*=52~\mathrm{ms}$, as field inhomogeneity in the sample causes the precessing nuclear spins to get out of phase with each other. A Fourier transform of this oscillating signal shows a sharp peak at the spin precession frequency (Fig.~\ref{fig:NmrSignalFFT}), with a signal-to-noise ratio of about 160. The width of this resonance divided by the drive frequency, 172.3 MHz, gives the inhomogeneity of the field in the NMR bulb, 24 ppb.  Fig.~\ref{fig:NmrSignalFFT} insert shows the Fourier transform, which has wider ``tails''.  This is not surprising given that the $^3$He gas in the glass capillary, just above the glass cell, contributes to the NMR signal and the magnetic field in the capillary is different than in the center of the solenoid field.  We thus concentrate upon the width of the central feature, shown in black line in the insert.  

\section{longitudinal and transverse relaxation time constants}
\label{sec:timeconstants}
There are three time constants that are important in NMR measurements, $T_1$, $T_2$, and $T_2^*$.  $T_1$ is the longitudinal relaxation time constant. $T_2$ is the decoherence time that would arise if the external magnetic field was perfectly homogeneous.  It is the effect of the fluctuating magnetic field of the spins upon each other and limits the linewidth of a NMR probe. $T_2^*$ is the NMR signal decoherence time that arises because of the magnetic field inhomogeneity in the magnetic field of the solenoid system. The best $T_2^*$ we have achieved is $52$ ms, as shown in Fig.~\ref{fig:NmrFID}. Here we discuss the measurement of $T_1$ and $T_2$.

\subsection{Measurement of longitudinal time constant $T_1$}

The time constant $T_1$ characterizes the time required for the initial thermal imbalance between the two spin states to be reestablished. Some measurements report $T_1$ of $^3$He to be as long as 1 day \cite{PhysRevLett.33.18, PhysRev.128.186, PhysRevA.48.4411, Lusher1988}, and we were initially worried that this time constant was so long that it might be difficult to make repetative measurements separated by only short times. 

We measure the $T_1$ time of our $^3$He sample by the saturation recovery method \cite{doi:10.1021/ed056p304}. First, a very long pulse drive compared to the $T_\pi$ is applied several times to randomize the spins of the $^3$He atoms. Then we wait for a certain length of time for the total magnetization of $^3$He to ``recover.'' Finally, apply a $\pi/2$ pulse to measure the magnitude of the NMR signal after the recovery time. The time evolution of the magnetization $M(t)$ follows
\begin{equation}
    M(t)=M_{0}\left[1-\exp\left(\frac{t}{T_1}\right)\right],
    \label{eq:saturationrecovery}
\end{equation}
where $M_0$ is the magnetization of the thermal $^3$He atoms, calculated from Eq.~\ref{eq:TotalMagneticMoment}.

\begin{figure}
    \centering
    \includegraphics[width=\the\columnwidth]{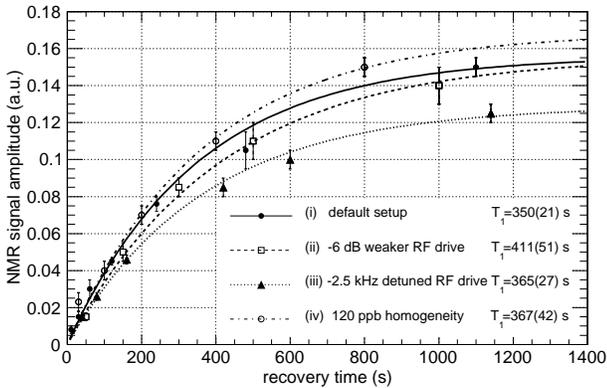}
    \caption{Measurement of the longitudinal relaxation time $T_1$ with saturation recovery method. The dots are measured data and the lines are fittings by Eq.~\ref{eq:saturationrecovery}. Several parameters are varied to check the consistency. See texts for details.}
    \label{fig:T1Measurement}
\end{figure}

The $T_1$ measurements with this method are shown in Fig.~\ref{fig:T1Measurement}. We also varied some parameters of the setup as systematic checks. In Fig.~\ref{fig:T1Measurement}, (i) is the default setup as in Figs.~\ref{fig:PiHalfMeasurement}, \ref{fig:NmrFID}, and \ref{fig:NmrSignalFFT}. As for the other measurements, (ii) a 6 dB RF attenuator is put in after the 26dB amplifier on the drive side, (iii) the RF drive frequency is 2.5 kHz detuned from the resonant frequency, and (iv) the z shim coil is intentionally ramped to make the homogeneity worse. All the measurements are consistent within their error bars. By assigning the largest discrepancy between them as the systematic error and taking the weighted mean of them, the longitudinal relaxation time is calculated to be $T_1 = 364~(31)$ s. The time constants from magnetic dipole interaction and diffusion are much longer than this measured value \cite{PhysRevA.48.4411, Lusher1988, PhysRevA.12.2333, PhysRevLett.33.18, PhysRevLett.34.1488, PhysRevA.37.2877}, and thus in our system, the wall relaxation effect is dominant. Similar results of $T_1$ measurements have been reported \cite{PhysRev.128.186, doi:10.1002/(SICI)1522-2594(199906)41:6<1084::AID-MRM2>3.0.CO;2-4, PhysRevA.83.061401}.

Even though $T_1$ time is long, it does not limit the application of our NMR probe. Due to the high signal-to-noise ratio achieved in our setup, as shown in Fig.~\ref{fig:NmrSignalFFT}, measurements with the recovery time of 20 seconds still give a signal-to-noise ratio of about 10. Usually we spend about this much time changing the current on shim coils to avoid the magnet quenching. A signal-to-noise ratio of 10 is good enough to see the effect of changing the currents on the shim coils. When we monitor the drift of the superconducting magnet, we usually take the NMR signal every 60 seconds. The drift rate of the magnet is much slower than the longitudinal time constant $T_1$, see also Sec.~\ref{sec:stability}.

\subsection{Measurement of transverse  time constant $T_2$}
$T_2$ is the relaxation time constant of the transverse magnetization even when the external magnetic field is perfectly homogeneous. This limits the coherence time of the NMR decay and thus sensitivity on the inhomogeneity of the magnetic field. Note that the $^3$He atoms are moving at an average speed of $v_\textrm{ave} = \sqrt{8k_BT/\pi m}=172$ m/s, and the relaxation timescale of $^3$He is on the order of 1 second. Therefore, even with the small diffusion coefficient, the effect is not negligible  here.

\begin{figure}
    \centering
    \includegraphics[width=\the\columnwidth]{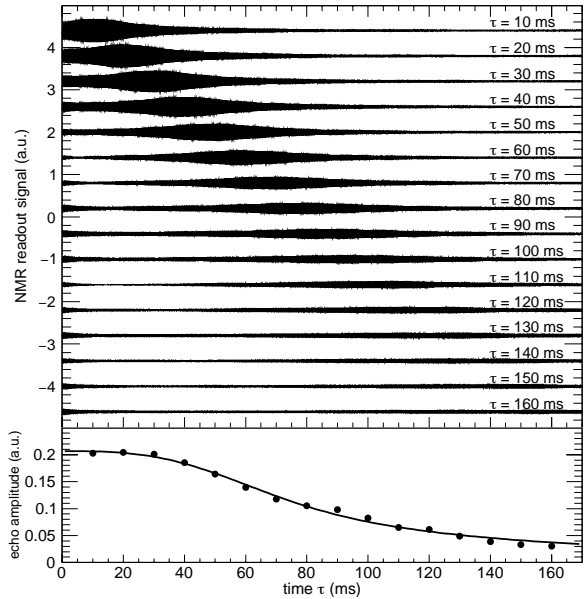}
    \caption{(top) Spin echo signals observed with the $^3$He NMR probe. Horizontal axis is the time after applying $\pi$ pulse. Measurements with 16 different echo times $\tau$ are shown. (bottom) Amplitudes of the echo are plotted as a function of echo time $\tau$. Solid line is a fitting by Eq.(\ref{eq:diffusionDecay}) without the $T_2$ term. See text for details.}
    \label{fig:SpinEchoSignal}
\end{figure}

\begin{figure}
    \centering
    \includegraphics[width=\the\columnwidth]{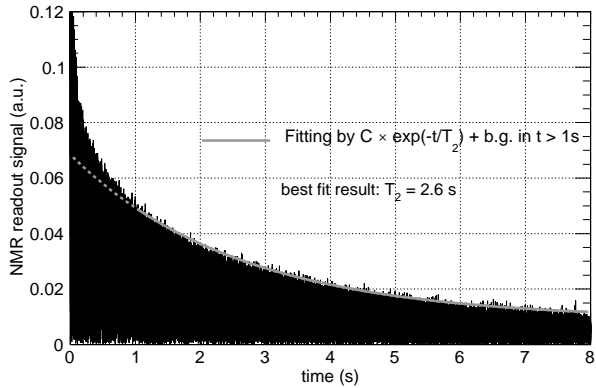}
    \caption{CPMG spin echo decay of $^3$He measured by the probe with the echo time $\tau = 5$ ms. The grey line shows an exponential fitting in $t>1$ s, which gives $T_2=2.6$ s as the best fit result. The fast decay signal at the beginning is due to diffusion of the $^3$He and quadratic magnetic field in our NMR volume \cite{BENDEL1990509, PhysRevB.46.3465}. See text for details.}
    \label{fig:CPMGEchoSignal}
\end{figure}

Figure \ref{fig:SpinEchoSignal} shows the spin echo measurements \cite{PhysRev.80.580} performed with the probe. In the graphs, $\pi/2$ pulse and $\pi$ pulses are applied at $t=-\tau$ and $t=0$ respectively. In the top graph, the interval between the $\pi/2$ pulse and the $\pi$ pulse, $\tau$, is varied among 16 different values. Clear echos corresponding to each $\tau$ are observed. The bottom graph shows the amplitudes of the echos as a function of $\tau$. As mentioned above, the effect of diffusion has to be taken into account to explain the observed decay of the echo signals. Since the data was taken after shimming the magnet, we assume the largest source of inhomogeneity is quadratic term of the magnetic field gradient. The amplitude of the first echo at $t=\tau$ as a function of the interval is given by \cite{BENDEL1990509, PhysRevB.46.3465}
\begin{eqnarray}
    A(\tau) &=& \frac{\sqrt{\pi}}{2\beta\tau^{1.5}}\mathrm{Erf}\left(\beta\tau^{1.5}\right)\exp(\frac{-2\tau}{T_2})    \label{eq:diffusionDecay}
\\
    \beta &=& \sqrt{\frac{8}{3}D\gamma^2b^2L^2},
\end{eqnarray}
where $b$ represents quadratic magnetic field gradient $B_z(\mathbf{r}) = B_0 + bz^2$, $D$ is the diffusion coefficient, $\gamma$ is the gyromagnetic ratio, $L$ is half of the typical size of the target volume, and $T_2$ is the intrinsic transverse decay constant. Fig.~\ref{fig:CPMGEchoSignal} shows the echo signal is fitted by Eq.~\ref{eq:diffusionDecay} assuming $\beta\tau^2 \gg 2/T_2$. The best fit result gives $\beta=0.0023$ (ms)$^\mathrm{-1.5}$.

The $\pm 2\sigma$ linewidth during these measurements is about 230 ppb. Based on the discussion of residual magnetism in Sec.~\ref{ch:magnetism}, we conservatively assume the uncertainty of this value to be $\pm50$ ppb. The typical straight line length of the bulb is $2L = \sqrt[3]{4\pi/3\times(0.5\mathrm{cm})^3}=0.8$ cm. Thus we estimate $b=\left(7.6 \pm 1.7\right)\times10^{-2}~\textrm{T/m}^2$. By using the gyromagnetic ratio of $^3$He, $\gamma= 2\pi\times32.434~\textrm{MHz/T}$, the diffusion coefficient is calculated to be $D = \left(5.9 \pm2.6\right)\times 10^{-7}$~m$^2$/s. This agrees with a previous measurement \cite{PhysRevA.39.6170} at a lower magnetic field that found $D=3.6\times 10^{-7}$~m$^2$/s. This small diffusion coefficient ensures that the motional narrowing effect is small in the $T_2^*$ measurements\cite{PhysRevA.37.2877}.

To minimize motional diffusion effects, we employ a Carr-Purcell-Meiboom-Gill (CPMG) pulse sequence \cite{PhysRev.94.630, doi:10.1063/1.1716296} and are thus able to measure $T_2$. As shown in Fig. \ref{fig:NMRCircuit}, in a CPMG echo measurement, multiple $\pi$ pulses are applied with $2\tau$ intervals, with the first $\pi$ pulse applied at $\tau$ after a $\pi/2$ pulse. The subscript $x$ and $y$ denote relative phases of the drive pulses. The interval between CPMG pulses is set to be $\tau=\tau_0=5$ ms. With this echo time, the effect of diffusion is negligible as long as $T_2 \ll 1/\beta^2\tau_0^2=7500$ s. Figure \ref{fig:CPMGEchoSignal} shows the CPMG spin echo signal. The initial rapid decay is a combined effect of diffusion and quadratic magnetic field \cite{BENDEL1990509, PhysRevB.46.3465}. The exponential decay at later times gives the time constant $T_2$. An exponential fitting in $t> 1$ s gives a transverse relaxation time constant $T_2=2.6$ s. A similar value has been obtained by a room temperature NMR measurement \cite{KOBER1999308}.

We also performed this measurement with different $\pi$ pulse lengths. The $\pi$ pulse length are varied by $\pm 3\%$, and the measured $T_2$ values fluctuates between 2.56 s and 2.81 s. We take this spread as a systematic error and estimate $T_2 = 2.7(2)~\textrm{s}$. The corresponding relative inhomogeneity is $1/(\omega_0T_2) = 0.34(3)$ ppb. This is much smaller than the linewidth we have achieved and thus does not limit the performance of our probe.

\section{Possible Magnetism of the Probe}
\label{ch:magnetism}

\begin{figure}
    \centering
    \includegraphics[width=\the\columnwidth]{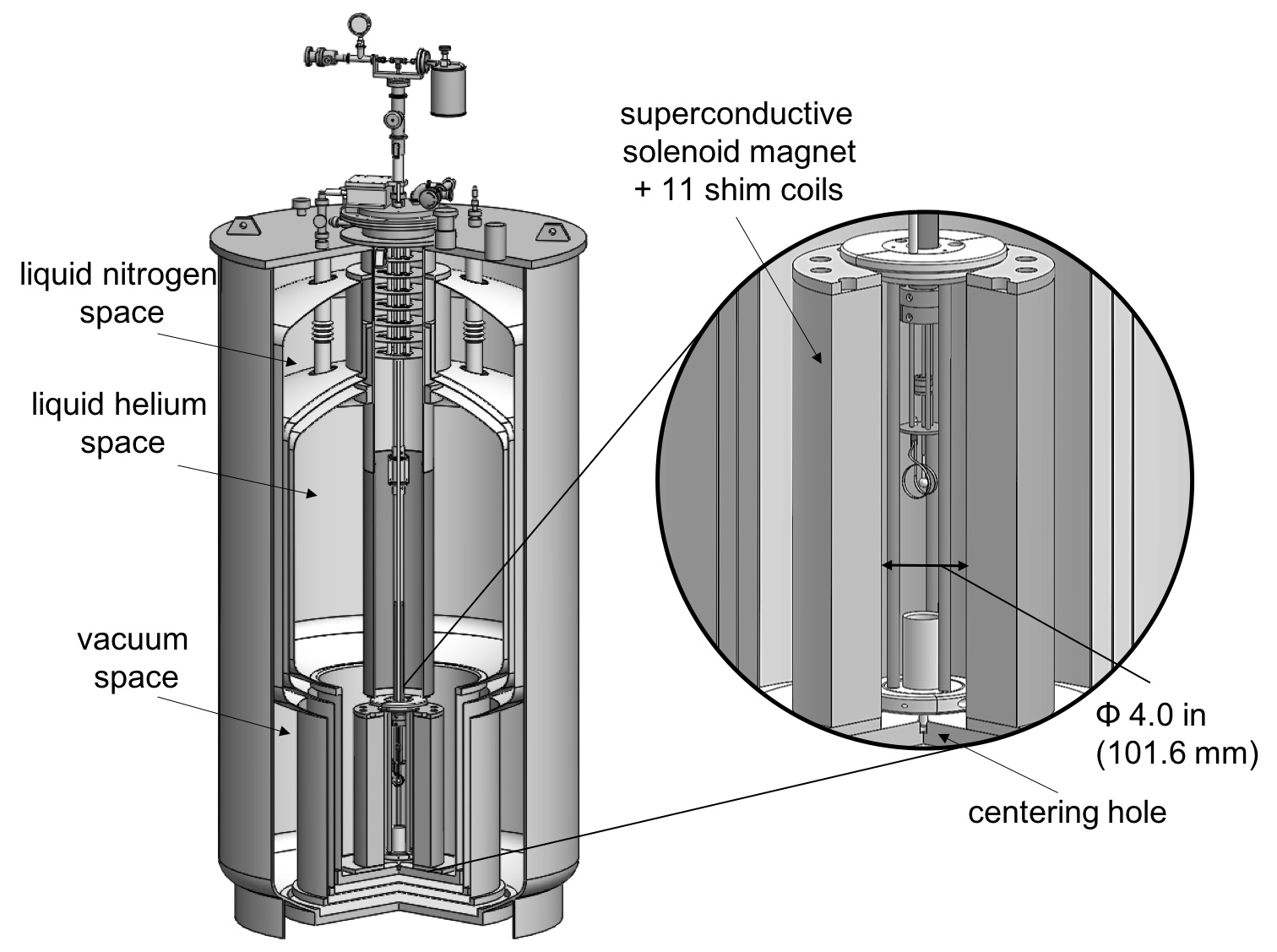}
    \caption{The superconducting solenoid system with the $^3$He NMR probe inserted into the 4.2 K cold bore. The bore size is 4.0 inches (101.6 mm) in diameter. It has 11 superconducting shim coils that can be used to optimize the homogeneity, in addition to the main 5.3 T solenoid magnet. The magnet is submerged in a liquid helium bath and thus the bore is always filled by liquid helium. NMR probe's centering plate and pin mate with the magnet structure as shown in the figure.}
    \label{fig:SolenoidSystem}
\end{figure}

In the end, a measurement that requires a high field homogeneity will need to have the magnetic field shimmed to take out the unavoidable magnetism of the measurement apparatus.  The magnetism of the NMR probe itself is one example. At the level of relative inhomogeneity we are interested in, $\mathcal{O}\left(10^{-9}\right)$, the magnetism of the probe is not necessarily negligible. In order to estimate the residual magnetism of the NMR probe itself, our NMR probe is designed so that the NMR bulb and its support structure can be rotated inside the magnet bore from the top of the dewar. Figure \ref{fig:SolenoidSystem} shows our superconducting solenoid system with the NMR probe inserted into the cold bore. The center rod that supports the NMR bulb is connected all the way to the top of the dewar. We can rotate the NMR bulb, capillary line and its supports, the electronics circuit board, and the RF coil. Note that the magnet mating parts in Fig.~\ref{fig:NmrProbeOverview} do not rotate, but they are made of pure copper and aluminum, and far from the bulb, while the rotatable parts are made of a variety of materials and some of them are quite close to the bulb. Thus the magnetism from the rotatable parts are estimated to be much higher than that from others. 

\begin{figure}
    \centering
    \includegraphics[width=\the\columnwidth]{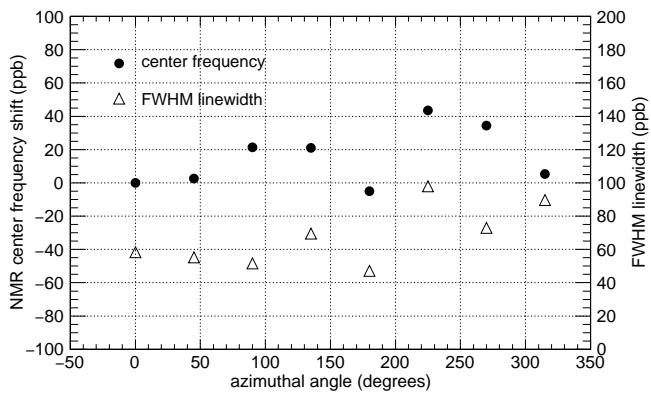}
    \caption{Dependence of the NMR center frequency and linewidth on the azimuthal angle of the probe. The initial orientation of the probe is defined to be 0 degrees. Both the center frequency and the linewidth depend on the angle of the NMR probe. The residual magnetism of the NMR probe can be estimated by the change of the center frequency and linewidth. See text for details.}  \label{fig:RotationMeasurement}
\end{figure}

After shimming the magnet, we rotate the probe to check the residual magnetism, the results of which are shown in Fig.~\ref{fig:RotationMeasurement}. The azimuthal angle of 0 degrees is defined as the initial position of the probe, and the initial center frequency is defined as 0 ppb shift on the left axis. Some dependence on the azimuthal angle can be seen. At the worst angle measured, 225 degrees, the linewidth increases up to about 100 ppb, while the center frequency changes by about 50 ppb. Thus the effect of the residual magnetism is estimated to be 50 ppb.

Possible candidates for the residual magnetism in the probe have been investigated. The absolute value of the magnetic field produced by a dipole magnetization vector $\mu$ at distance $r$ is $|B|\sim{\mu_0}/{4\pi}\times{\mu}/{r^3}$, where $\mu_0$ is the vacuum permeability. The value varies by a factor of 2 at most depending on the direction of the magnetization vector. As can be seen, the effect from residual magnetism is proportional to $r^3$. The closest part has by far the largest contribution to the magnetic field inhomogeneity. 

In our case, the closest part is the RF copper coil. The magnetism of the 99.9999$\%$ purity copper foil that is used in the NMR probe is measured by a SQUID magnetometer (MPMS 3, Quantum Design Inc. \cite{squidmagnetometer}) to be $\left(5.0 \pm 1.2\right) \times 10^{-5}$ J/T~cm$^3$ at 5.3 T. Our coil is made of a foil of cross section 0.1~mm$\times$3~mm. Even with this small volume an inhomgeneity would be seen. For example, if 1 cm of this foil is placed at 5 mm away from the NMR bulb, it will induce about 40 ppb inhomogeneity. This is as large as the rotational dependence we have observed. Searches for high Q-factor conductive metals with lower magnetism to replace the current copper coil are underway.
Measurements using a small fragment of all other materials used (e.g. copper, tungsten, aluminum, circuit board, capacitor, SMA connector, glass of the bulb) suggest that these give smaller contributions. 

\section{Stability of a Cold Bore Solenoid System}
\label{sec:stability}
The superconducting solenoid used for these studies (Fig.~\ref{fig:SolenoidSystem}) was custom designed for positron and electron magnetic moment measurements that will be carried out to test the most precise prediction of the standard model of particle physics \cite{TowardsElectronPositronMoments2019}. For the lepton measurements, a Penning trap apparatus and a dilution refrigerator are lowered into the central bore of the superconducting solenoid system. For the measurements reported here, the NMR probe apparatus is inserted instead of the dilution refrigerator so that the bulb containing the $^3$He gas is at the location that a single lepton would be suspended. 

The top of the metal form on which the solenoid is wound has a register that is used to center the NMR probe and the lepton trap apparatus, and a lower hole that allows a centering pin to center the inserted apparatus at the bottom of the solenoid.  The lepton apparatus is set directly upon the 4.2 K solenoid form so there can be no relative motion of the magnetic field and the apparatus making use of it.

The lepton measurements are made using  quantum jump spectroscopy measuring the lepton spin and cyclotron frequencies.  To achieve a smaller uncertainty the cyclotron frequency and the difference of the spin and cyclotron frequencies (the anomaly frequency) are actually measured.  All of these frequencies are proportional to magnetic field, and this field dependence cancels out to lowest order since the magnetic moment depends on the ratio of these two frequencies.  However, field variations are bad insofar as they increase the measured linewidths, and as the two frequencies are measured at slightly different times, separated by about 1 minute. To determine the electron magnetic moment to the current accuracy of $3$ parts in $10^{13}$, we have to perform spectroscopy at $3$ parts in $10^{10}$. For example, if we want to improve the limit by a factor of 10, the field needs be stable or corrected for drifts better than 1.8 ppb/h.

\begin{figure}
    \centering
    \includegraphics[width=\the\columnwidth]{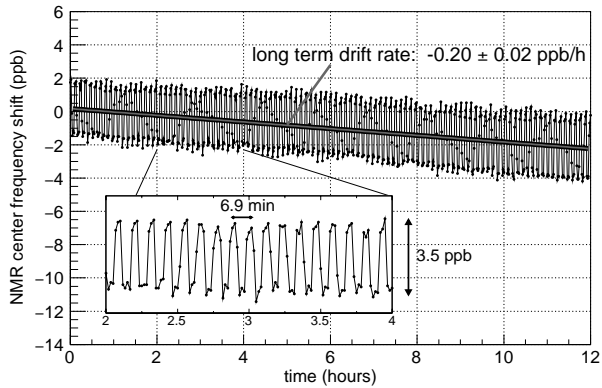}
    \caption{Measurement of the $^3$He center frequency drift. $^3$He NMR signal is taken every minute and fitted by a Lorentzian to acquire center frequency. The long term drift rate is good enough to perform future magnetic moments measurements. The probe also reveals periodic oscillations that have not previously been observed. See text for details.} 
    \label{fig:DriftMeasurement}
\end{figure}

Once a narrow NMR linewidth is determined, we measure this frequency repetitively to determine the field stability.  We are interested in the stability on times over which we make individual frequency measurements, and the long term stability over the time that it takes to make the number of frequency measurements that must be averaged to get the desired uncertainty.  
The uncertainty in the NMR spin precession frequency in a single measurement is about 0.1 Hz, which corresponds to 0.5 ppb. By averaging the measurements over a long period, we can reduce the uncertainty on the drift rate.  

Fig.~\ref{fig:DriftMeasurement} shows a sequence of measurements performed. The $^3$He NMR signal is taken every minute and is fitted by a Lorentzian to get the center frequency. The center frequency has been monitored for 12 hours, which shows a drift of about -0.2 ppb/h, but it also revealed a surprising periodic and non-negligible oscillation. The average pressure in the helium space and the nitrogen space of the solenoid system were well regulated during this measurement. There are no obvious correlations with monitored temperatures, pressures, and flow rates in the helium and nitrogen dewars or ambient magnetic field, temperature, humidity, and atmospheric pressure in the lab. The drive and mixdown power and frequency were also monitored, but no obvious correlations were found. Studies, enabled by the gaseous NMR probe, are underway to understand and eliminate these oscillations. 

\section{conclusion}
A gaseous $^3$He NMR probe at 4.2 K has a signal-to-noise and linewidth comparable to room temperature probe with the same volume that uses a liquid that is rich in hydrogen, such as water.  A coupled reservoir-bulb system is key to attaining this signal-to-noise without requiring a dangerously high gas pressure. Several measurements that are intrinsic to an NMR spectroscopy have been performed. We show that the gas probe is ideal for shimming and measuring the stability of superconducting solenoids with cryogenic bores.     

\section{acknowledgements}
J. Dorr, E.\ Novitski, and M. Dembecki contributed to an earlier version of this probe.  This work was supported by the NSF, with X. Fan being partially supported by Masason Foundation.

\bibliographystyle{prsty_gg}
\bibliography{output-2.bbl}

\end{document}